\documentclass[twocolumn,amsmath,amssymb,preprintnumbers,showpacs,pra]{revtex4}

\usepackage{graphicx,color}

\usepackage{amsmath}
\usepackage{amssymb}
\usepackage{amsfonts}
\usepackage{amsthm}

\newtheorem{lemma}{Lemma}

\begin{document}

\title{Discrete Rotational Symmetry and Quantum Key Distribution Protocols}

\author{David Shirokoff}%
\affiliation{%
}%

\author{Chi-Hang Fred Fung}%

\author{Hoi-Kwong Lo}
\affiliation{%
Center for Quantum Information and Quantum Control,\\
Department of Electrical \& Computer Engineering and Department of Physics,\\
University of Toronto, Toronto,  Ontario, Canada\\
}%


\begin{abstract}
We study the role of discrete rotational symmetry in quantum key
distribution by generalizing the well-known Bennett-Brassard 1984
(BB84) and Scarani-Acin-Ribordy-Gisin 2004 (SARG04) protocols. We
observe that discrete rotational symmetry results in the protocol's
invariance to continuous
rotations, thus leading to
a simplified
relation between bit and phase error rates and consequently a
straightforward security proof.

\end{abstract}

\maketitle

\section{Introduction}

Unlike classical cryptography which often relies on unproven
assumptions on computational complexity to ensure secrecy, quantum
cryptography, or more precisely, quantum key distribution (QKD),
provides proven unconditional security against any attacks allowed
by quantum mechanics. To date, most of these security proofs handle
protocols on a case by case basis. In particular, two well-known QKD
protocols, the Bennett-Brassard 1984 (BB84) protocol
\cite{BB84,Mayers,LC,BBBMR,SP,Ekert1991} and the
Scarani-Acin-Ribordy-Gisin 2004 (SARG04) protocol
\cite{SARG04,BGKS,TL,FTL}, have a collection of independent proofs
despite sharing a variety of properties.  For example, both schemes
are {\it prepare-and-measure} protocols, implying that
implementation does not require a quantum computer but rather is
within the ability of modern optical technology.
Both protocols
utilize qubits and
use the same four quantum states to encode information. Thus, the
quantum transmission part is the same. They differ only in the classical
discussion
part (in particular, the basis reconciliation step).
The two protocols organize the quantum states into bases in different manners;
however, both
exhibit discrete rotational symmetry among the
states.
The fundamental difference
between them arises from the fact that the pairs of basis states in
SARG04 are not orthogonal as they are in BB84.  This
non-orthogonality makes the SARG04 protocol more resilient to
multiple-photon attacks and thus plays a practical role.

In this letter, we wish to generalize the BB84 and SARG04 protocols to allow an arbitrary number of
bases ($M$) and arbitrary angle ($\theta$) between basis pairs. Our motivation is to better
understand the role of discrete symmetry in QKD by studying the relation between these parameters
and the resulting protocol security. The previous work of Koashi \cite{MK} has made strides in the
subject. Nonetheless, we improve on his results in several areas. Firstly, we derive a closed form
formula for the relationship between bit and phase error rates, thus allowing us to place tighter
bounds on the quantum key generation rate. Secondly, our results hold for any value of $\theta$,
while Koashi requires $\theta$ to take on discrete values related to $M$. Lastly, rather
surprisingly, our closed form expression is independent of the number of the bases $M$ for $M > 2$.
 Note that our result does not subsume Koashi's, since he studied the relations between bit and
phase error rates for an arbitrary number of photons sent out by Alice
while we consider only the single-photon case.

\section{The generalized protocol}
\begin{figure}[h]
   \centering
   \title{Basis States for $M = 4$}
      \includegraphics[width = .9\columnwidth]{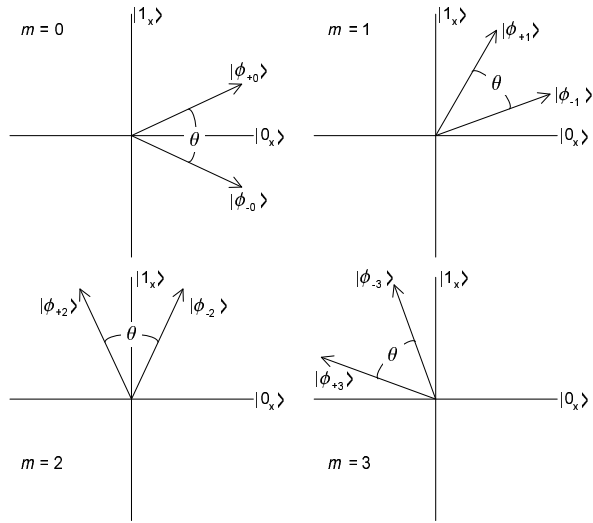}
   \caption[The basis states for the generalized protocol]{The four graphs show the different pairs of qubits created by Alice ($0 \leq m \leq 3$). We allow for an
   arbitrary angle between the basis pairs $\theta$. In the case when $M = 4$, $\theta = \pi/2$, we recover BB84 (a symmetrized version with four bases), while $\theta = \pi/4$ recovers SARG04. One should note that when $\theta = \pi/2$, the states $\phi_{\pm
m}$ become z-eigenstates.}
   \label{fig:MBasis}
\end{figure}

We now describe the
generalized QKD protocol. In this protocol, Alice prepares qubit
states. Experimentally, these may be realized by polarization
states. In the SARG04 protocol, Alice chooses from one of four pairs
of basis states. We generalized the protocol, so that now she
chooses from a set of $M \geq 2$ pairs. Hence each of the $M$ basis
pairs has a $1/M$ chance of being selected. Moreover, for each of
the $M$ basis pairs, the two states are separated by an angle
$\theta$ $>$ 0 (see Fig.~\ref{fig:MBasis}).
Thus, when $M = 4, \theta =
\pi/2$, the collection of states is identical to BB84 (here, we
adopt a symmetrized version of BB84 with four bases instead of two),
while $M = 4, \theta = \pi/4$ reduces to the SARG04 protocol.
Note that our generalized protocol does not include the six-state
protocol \cite{Bruss,Inamori,Lo} since the states in our generalized
protocol lie on a plane in the Bloch sphere whereas the six states
in the six-state protocol do not.

In our notation it is convenient to introduce the rotation operator about the y-axis
$\widehat{R}_{\beta}=e^{-\imath \beta \widehat{\sigma}_y}$, where $\widehat{\sigma}_{y} =
\imath(|1_x\rangle\langle0_x|-|0_x\rangle\langle1_x|)$. Here, the coefficient of $1$ as opposed to
$1/2$ in the exponential arises from the fact that we are considering a spin-$\pm 1$ qubit, such as
a photon, embedded in a 3D Hilbert space. Thus, the x-basis kets satisfy: $|1_x\rangle =
\widehat{R}_{\pi/2}|0_x\rangle$.

We now denote the set of states that Alice may choose from by: $\{|\phi_{+m}\rangle,
|\phi_{-m}\rangle\}$ for $0 \leq m \leq M-1$. Furthermore, these states take the following
representation:
\begin{eqnarray}\label{rotation}
     |\phi_{\pm m}\rangle 
     &=& \widehat{R}_{m\pi/M} \left(\cos\frac{\theta}{2} |0_x\rangle \pm \sin\frac{\theta}{2} |1_x\rangle\right)
\end{eqnarray}
where $|0_x\rangle$ and $|1_x\rangle$ are x-basis eigenkets. Thus, when $\theta = \pi/2$, the
states $|\phi_{\pm m}\rangle$ become z-basis eigenkets.  Also, by construction, the set of basis
states $|\phi_{\pm m}\rangle$ are simply rotated copies of each other, equally spaced at angular
intervals of $\pi/M$, about the upper half unit circle.  Due to the rigid rotation of the basis
states, each pair of states satisfies the inner product $\langle\phi_{+m}|\phi_{-m}\rangle =
\cos\theta$, for $0 \leq m \leq M-1$.

We outline the generalized prepare-and-measure protocol as follows:
\begin{description}

  \item[1] Alice prepares n qubits. For each qubit, she
  randomly chooses one of the $M$ bases
  pairs.  We denote her choice of basis by the set of numbers:
  $0 \leq l_{i} \leq M-1$, where $1 \leq i \leq n$.
  For each qubit, she also picks a random bit $b_i$.
  \item[2] For the $i$th state, Alice chooses the $l_i$ basis and prepares
  either the state
  $|\phi_{l_i}\rangle$ if $b_i= 0$ or the state $|\phi_{-l_i}\rangle$ if
  $b_i=1$.
  \item[3] Alice sends her states over a quantum channel to Bob.
  \item[4] Bob receives $n$ possibly noisy qubits. For each state $i$, Bob chooses a random number
  $k_i$ representing the $k$th basis.
  For each $k_i$, he then randomly chooses one of two projection measurements
  (one projects onto the orthogonal states $\{|\phi_{k_i}\rangle, \widehat{R}_{\pi/2}|\phi_{k_i}\rangle\}$ and the other onto $\{|\phi_{-k_i}\rangle, \widehat{R}_{\pi/2}|\phi_{-k_i}\rangle\}$)
  to measure the incoming state.
  When the measurement outcome is $\widehat{R}_{\pi/2}|\phi_{k_i}\rangle$ or $\widehat{R}_{\pi/2}|\phi_{-k_i}\rangle$,
  he concludes that the incoming state is $|\phi_{-k_i}\rangle$ or $|\phi_{k_i}\rangle$, respectively.
  When either of the other two outcomes occurs, he declares an inconclusive result for this state.
  \item[5] Alice and Bob discuss over classical channels their preparation and measurement basis. They
  discard all $i$ states where $l_i \neq k_i$.  Furthermore, the pair discard all states where Bob
  records an inconclusive measurement.
\end{description}
  Hence, on average, Alice and Bob retain $N_{con} \approx n \cdot
  p_{con}/M$ bits, where $p_{con}$ is the probability of Bob's measurement is  conclusive.
  When the quantum channel is perfect (i.e., no Eve), $p_{con}=\frac{1}{2} \sin^2 \theta$.
  For some special cases such as the SARG04 protocol, the reduction by a factor of $M$ can be avoided (i.e., $N_{con} \approx n \cdot p_{con}$).
  This is because the two projection measurements for the different $k_i$'s are the same for SARG04.
  In other words, there are only two distinct measurements and are independent of $k_i$, but the interpretation of the measurement outcomes is dependent on $k_i$.
  Additionally, by the same token, the reduction by a factor of $2$ in $p_{con}$ may not be needed in some protocols such as the BB84 protocol.
  This is because the two projection measurements for each $k_i$ are the same for BB84.

Due to the possibility of a noisy quantum channel or the presence of Eve, the sequence of $N_{con}$ bits
shared by Alice and Bob will not be identical but rather have some fraction, $e_b$, of errors.
Also, Eve may have some knowledge of the $N_{con}$ bits.
The remaining steps in the protocol describe the process of converting the insecure sequence of
$N_{con}$ bits into a shorter sequence of $R\cdot N_{con} \cdot (1-t)$ secure bits. These steps are analogous
to those in BB84:

\begin{description}
  \item[6] Alice and Bob discuss and randomly choose a fraction $t$ of their $N_{con}$ bits. They reveal the
  values of their bits and estimate the error rate.
  \item[7] In a large sampling, their estimated error rate will asymptotically approach the true
  error rate $e_b$.
  \item[8] Depending on the value of $e_b$, the pair either abort the protocol or perform error correction and
  privacy amplification to distill a secure key of length $R\cdot N_{con} \cdot(1-t)$.
\end{description}

Without loss of generality and for simplicity, the states lost in the channel are excluded in the
description above and also in the following derivations. To facilitate the study, we first convert
our prepare-and-measure protocol into an equivalent entanglement-distillation-protocol (EDP)-based
QKD protocol (see \cite{SP}). By the equivalence of the two protocol formulations, any results we
derive with one, such as the bound on the secret key generation rate, apply directly to the other.

To appreciate the connection between EDP and QKD, we observe that if Alice and Bob shared $n$ pure
EPR states of the form $|\psi\rangle = \frac{1}{\sqrt{2}}(|00\rangle + |11\rangle)_z$, then
individual measurements in the basis $|0_z\rangle$, $|1_z\rangle$ will produce $n$ random,
identical results. That is, Alice and Bob share an identical string of random states composed of
either $|0_z\rangle$ or $|1_z\rangle$. After associating logical values with the states, Alice and
Bob will share a secure sequence of $n$ random digits, exactly what is required for a key.

In an EDP, Alice and Bob share $n$ entangled bipartite quantum states. These states, are either
mixed or pure. The task of generating an encryption key therefore reduces to the problem of
converting the entangled states into pure EPR pairs by using only local operations and classical
communications (LOCCs). Once the couple obtains pure EPR pairs, they may simply measure their
states and obtain a secure key as outlined above.

To convert our generalized protocol into an EDP-based protocol, we must determine a suitable
mathematical representation for the bipartite quantum states, $\rho_{AB}$, shared by Alice and Bob.
We observe that in the prepare-and-measure scheme, if Alice chooses the $l$th basis and randomly
prepares either the state $|\phi_l\rangle$ or $|\phi_{-l}\rangle$, then mathematically her
preparation process is identical to the construction of the joint quantum state
$|\psi_l\rangle_{AB} = \widehat{I}\otimes\widehat{R}_{l\pi/M}|\psi_0\rangle_{AB} =
\frac{1}{\sqrt{2}} (|0_z\rangle_A |\phi_l\rangle_B + |1_z\rangle_A |\phi_{-l}\rangle_B)$, followed
by a z-eigenbasis measurement in the $A$ quantum subspace.  The mathematical representation for
$\rho_{AB}$ is outlined as follows. In the subsequent notation we denote $\textbf{P}\Big(
|\cdot\rangle\Big) = |\cdot\rangle\langle\cdot|$.

\begin{description}
  \item[Alice's Preparation]:
  Alice prepares her states by choosing, with equal probability, among the different $M$
  bases. Hence in the EDP she will prepare the ensemble $|\psi_l\rangle$, for $0 \leq l \leq M-1$.
  Thus, for each $l$, Alice constructs the state:
  \begin{equation}
   \rho^l_{AB}= \textbf{P}\Big(\widehat{I}_A\otimes(\widehat{R}_{l\pi/M})_B
   |\psi_0\rangle_{AB}\Big).
    \end{equation}
  \item[Noisy Channel]:
  Alice sends the second qubit across a noisy channel to Bob.
  To model the noisy channel we
  introduce a general quantum operator $\mathcal{E}$, which maps density matrices to density matrices. Such an
operator accounts for all possible quantum processes including, unitary evolution in an extended
Hilbert space and generalized POVM measurements made by an eavesdropper.
In general, $\mathcal{E}(\rho)$ has a representation by a finite collection of operators $\{
\widehat{E}_j\}$ such that each $\widehat{E}_j \widehat{E}_j^\dag$ is positive, $\sum \widehat{E}_j
\widehat{E}_j^\dag = \widehat{I}$, and $\mathcal{E}(\rho) \rightarrow \sum \widehat{E}_j \rho
\widehat{E}_j^\dag$. Thus, the noisy quantum channel transforms the density matrix according to:
    \begin{eqnarray}
  \rho^l_{AB}&\rightarrow& \mathcal{E}(\rho^l_{AB})   \\
   &=& \sum_{j}\textbf{P}\Big(\widehat{I}_A
  \otimes (\widehat{E}_j\widehat{R}_{l\pi/M})_B|\psi_0\rangle_{AB}\Big). \nonumber
    \end{eqnarray}
  \item[Bob's Filtering]:
Lastly, when Bob receives his quantum states, he randomly chooses one of the $M$ bases to perform
his measurements. For each choice, he further selects one of two projection measurements to measure
the quantum state. As in Ref.~\cite{TL}, this measurement process can be modeled by a filtration
operation $\widehat{F}$, followed by a regular orthogonal measurement in the z-eigenbasis. Here,
the Kraus filtration operator for the pair of states $|\phi_{+0}\rangle$ and $|\phi_{-0}\rangle$
is:
$\widehat{F}_0 = \sin(\theta/2)|0_x\rangle\langle0_x|+\cos(\theta/2)|1_x\rangle\langle 1_x|$. The
corresponding filtration operator for a general pair of states $|\phi_{-l}\rangle$ and
$|\phi_{+l}\rangle$ is the rotated filter: $\widehat{F}_l = \widehat{F}_0 \widehat{R}_{-l\pi/M}$.
The Kraus filtration operator only transforms on the right side to ensure the invariance of
$\widehat{F}_0 |\psi_0\rangle = \widehat{F}_l |\psi_l\rangle$. Since Alice and Bob only retain
states in which they choose the same preparation and measurement basis, Bob's operation in the EDP
corresponds to simply $\widehat{F}_l = \widehat{F}_0 \widehat{R}_{-l\pi/M}$. Hence we have:
\begin{equation}\label{AB_densitymatrix}
       \rho_{AB} = \frac{1}{M}\sum_{j}\sum_{l = 0}^{M-1} \textbf{P}\Big(\widehat{I}_A
  \otimes (\widehat{F}_0\widehat{R}_{-l\pi/M}\widehat{E}_j\widehat{R}_{l\pi/M})_B|\psi_0\rangle_{AB}\Big).
\end{equation}
\end{description}

To sum over the $M$ bases in equation (\ref{AB_densitymatrix}), we expand out $\widehat{E}_j =
\widehat{U}_j + \widehat{V}_j$, with $\widehat{U}_j = a_i^j\widehat{\sigma}_i +
a_y^j\widehat{\sigma}_y$, $\widehat{V}_j = a_x^j\widehat{\sigma}_x + a_z^j\widehat{\sigma}_z$ and
complex coefficients $a_r^j$ for $r = i$, $x$, $y$, $z$. Here $\widehat{\sigma}_r$ are
representations of the identity or Pauli matrices. Thus, $\widehat{V}_j$ anti-commutes with
$\widehat{\sigma}_y$ while $\widehat{U}_j$ commutes with both $\widehat{\sigma}_y$ and
$\widehat{R}_\beta$. It is also simple to prove from first principles that for $M > 1$,  $\sum_{l =
0}^{M-1}\widehat{R}_{2l\pi/M} = 0$. Therefore we obtain:
\begin{eqnarray}\label{NewDensityMatrix}
       \rho_{AB} & = & \sum_{j} \bigg[\textbf{P}\Big((\widehat{F}_0\widehat{U}_j)_B|\psi_0\rangle_{AB}\Big)+
                \nonumber \\
                & & \frac{1}{M}\sum_{l = 0}^{M-1}
                \textbf{P}\Big((\widehat{F}_0\widehat{V}_j\widehat{R}_{2l\pi/M})_B|\psi_0\rangle_{AB}\Big)\bigg]\\
                & = & \widehat{F}_0 \sum_{j} \bigg[\widehat{U}_j|\psi_0\rangle_{AB} \langle\psi_0|\widehat{U}_j^\dag + \widehat{V}_j  \Phi  \widehat{V}_j^\dag
                \bigg]\widehat{F}_0^\dag.
\end{eqnarray}
where $\Phi = \frac{1}{M}\sum_{l = 0}^{M-1}
                \widehat{R}_{2l\pi/M}|\psi_0\rangle_{AB}\langle\psi_0|\widehat{R}_{-2l\pi/M} = \frac{1}{2}(\widehat{I}_A\otimes\widehat{I}_B+ |0\rangle_{z}\langle1|\otimes
\widehat{R}_{-\theta} +|1\rangle_{z}\langle0|\otimes \widehat{R}_{\theta})$. Here the last equality holds only for $M > 2$ while in the case of $M = 2$, $\Phi = |\psi_0\rangle_{AB}\langle\psi_0|$.

What is the effect of discrete averaging on $\rho_{AB}$?  Firstly, we observe that an arbitrary
continuous rotation through an angle $\beta$ by Alice and Bob transforms $\widehat{E}_j \rightarrow
\widehat{R}_{-\beta} \widehat{E}_j \widehat{R}_{\beta} = \widehat{U}_j +
\widehat{V}_j\widehat{R}_{2\beta}$.  Moreover, a close inspection shows that for $M > 2$, $\Phi$ commutes with
$\widehat{I}\otimes \widehat{R}_\beta$.  Therefore, the process of discrete averaging results in
the invariance of $\Phi$ and $\rho_{AB}$ to continuous rotations.  The origin of the symmetry
derives from the mathematical form of $\Phi$. By construction $\Phi$ commutes with Bob's discrete
rotations of $2l\pi/M$ and therefore has the same eigenspace as the discrete rotation operators.
For $M > 2$, discrete and continuous rotations share the same eigenspace.  Therefore $\Phi$ must
commute with all continuous rotations \cite{NoteB}.  In contrast, when $M = 2$, the rotation operators under consideration are proportional to the identity.  Hence, in this case, $\Phi$ does not have the same eigenspace as the continuous rotation group and so the general $M = 2$ protocol is not invariant under simultaneous rotations by Alice and Bob.

\section{Security analysis}

We now proceed with the security analysis. The quantum bit error rate, $e_b$, is the physical error
rate experienced by Alice and Bob during QKD.  Such an error occurs when Alice prepares a logical
$0$ or $1$, while Bob conclusively measures $1$ or $0$. The bit error only gives a partial
indication of the properties of a quantum channel.  One can imagine an eavesdropper or noisy
channel altering the relative phase of a qubit without effecting bit error rate statistics.
 Just as $e_b$ was introduced as the error obtained from Alice and Bob's preparation and measurement
in the z-basis, we may introduce the phase error as the resulting error rate if the couple prepared
and measured in the x-basis. To study bit and phase errors, we introduce the Bell basis:
\begin{eqnarray}\label{bellstates}
     |\Phi^\pm\rangle = \frac{1}{\sqrt{2}}(|00\rangle \pm |11\rangle)_z,
     |\Psi^\pm\rangle = \frac{1}{\sqrt{2}}(|01\rangle \pm |10\rangle)_z.
\end{eqnarray}
The Bell diagonal elements of Alice and Bob's bipartite density matrix represent the probabilities
of a quantum channel inducing various errors on their shared entangled state. That is, we may think
of Alice and Bob originally sharing an entangled EPR pair while the effect of an eavesdropper or
noisy channel is to induce bit errors $|0\rangle_z \rightarrow |1\rangle_z$ and vice versa, or
phase errors $\frac{1}{\sqrt{2}}(|0_z\rangle \pm |1_z\rangle) \rightarrow
\frac{1}{\sqrt{2}}(|0_z\rangle \mp |1_z\rangle)$. Thus, the diagonal elements of the bipartite
density matrix have the following form:
\begin{eqnarray}\label{bellprobabilities}
     p_I = \langle\Phi^+|\rho_{AB} |\Phi^+\rangle,& &
     p_x = \langle\Psi^+|\rho_{AB}|\Psi^+\rangle\\
     p_y = \langle\Psi^-|\rho_{AB}|\Psi^-\rangle,& \phantom{xx}&
     p_z = \langle\Phi^-|\rho_{AB}|\Phi^-\rangle .
\end{eqnarray}

Here, $p_x$, $p_z$ and $p_y$ represent the probabilities of a bit flip, phase flip and both bit and
phase flip error.  From the diagonal elements, we may extract closed form expressions for the
errors, since $e_b =  p_x + p_y$ and $e_p = p_y + p_z$.
We may
insert matrix representations and
calculate the diagonal elements of $\rho_{AB}$ explicitly.
[As stated earlier, we always assume that $\theta>0$.]
Interestingly, in
the case when $M > 2$,
Eq.~(\ref{AB_densitymatrix}) simplifies significantly because of spherical averaging, resulting in expressions that are independent of $M$;
on the other hand, in
the case when $M = 2$, spherical averaging does not simplify
Eq.~(\ref{AB_densitymatrix}).

\subsection{Case one: $M>2$}
For the $M>2$ case, we have
\begin{eqnarray}
     p_i &=& \frac{1}{N'} \sum_j 2|a_i^j|^2\sin^2\theta \label{pi_calculations}\\
     p_x &=& \frac{1}{N'} \sum_j \sin^2\theta (|a_x^j|^2 + |a_z^j|^2)\label{px_calculations}\\
     p_y &=& \frac{1}{N'} \sum_j (2|a_y^j|^2+\cos^2\theta (|a_x^j|^2 + |a_z^j|^2) )\\
     p_z &=& \frac{1}{N'} \sum_j (|a_x^j|^2 + |a_z^j|^2 + 2\cos^2\theta |a_y^j|^2) \label{pz_calculations}
\end{eqnarray}
where $N'$ is a normalization constant (see Appendix~\ref{app-caldiagonal} for detail).  
Thus, the
relation between the bit and phase error rate is
\begin{equation}\label{bitphaseerror}
    e_p = e_b(1+\cos^2 \theta), \phantom{xxxxxxxx} \text{for } M>2.
\end{equation}

We remark that this relation as well as the diagonal elements of the
marginal density matrix, Eqs. (10)-(13), hold independent of the
loss in the channel, and also independent of the number of bases $M$
provided $M > 2$. In fact the only important value that still
depends on $M$ is the ratio of conclusively measured bits,
$N_{con}$, to the total signal length $n$: $N_{con}/n$. Here the
ratio scales as $1/M$, since Bob has a $1/M$ probability of choosing
the same basis for his measurements as the one used by Alice during
preparation. Consequently, there is little practical advantage for
increasing the number of bases in the generalized 2D protocol.

\subsection{Case two: $M=2$}

For the $M=2$ case, the diagonal elements of $\rho_{AB}$ are
\begin{eqnarray}
     p_i &=& \frac{1}{N''} \sum_j |a_i^j|^2\sin^2\theta \\
     p_x &=& \frac{1}{N''} \sum_j |a_x^j|^2\sin^2\theta \\
     p_y &=& \frac{1}{N''} \sum_j (|a_y^j|^2 + |a_z^j|^2 \cos^2\theta)\\
     p_z &=& \frac{1}{N''} \sum_j (|a_z^j|^2 + |a_y^j|^2 \cos^2\theta),
\end{eqnarray}
and the bit and phase error rates are
\begin{eqnarray}
     \label{eqn-M2eb}
     e_b &=& \frac{1}{N''} \sum_j (|a_x^j|^2\sin^2\theta + |a_y^j|^2 + |a_z^j|^2 \cos^2\theta)\\
     \label{eqn-M2ep}
     e_p &=& \frac{1}{N''} \sum_j (|a_y^j|^2 + |a_z^j|^2)(1 + \cos^2\theta),
\end{eqnarray}
where $N''=\sum_j \sin^2\theta(|a_i^j|^2+|a_x^j|^2)+(1+\cos^2\theta)(|a_y^j|^2+|a_z^j|^2)$.
Note that a direct relation between the bit and phase error rates cannot be readily observed for this case.
In order to obtain this relation, we solve the problem of maximizing the phase error rate
subject to a fixed bit error rate (see Appendix~\ref{app-epebrelation} for detail).
The resulting relation between bit and phase error rates is as follows:
\begin{equation}
\label{bitphaseerror2}
     e_p \leq \frac{1+\cos^2 \theta}{\cos^2 \theta} e_b, \phantom{xxx} \text{for }
     M=2 .
\end{equation}
Note that when $\theta=\pi/2$, the phase error rate $e_p$ goes to infinity. 
This is because the two states in the second basis of the protocol are the same (except for the phases) as the two states in the first basis (cf. Fig.~\ref{fig:MBasis}).
Thus, the protocol consists of only two orthogonal states and QKD cannot be performed.

\subsection{Key generation rate}
We may utilize Shor-Preskill's argument \cite{SP} to obtain the
secret key generation rate \cite{NoteA} as (assuming the fraction
$t$ of test bits goes to zero)
\begin{equation}
\label{eqn:keyrate1}
R_{\text{final}} =  \frac{p_{con}}{M} [1-H_2(e_b)-H_2(e_p)],
\end{equation}
where $H_2(p)=-p \log_2(p) - (1-p) \log_2(1-p)$ is the binary
entropy function, $e_b$ is the observed bit error rate, and $e_p$ is
the inferred phase error rate from Eq.~\eqref{bitphaseerror} (for
$M>2$) or Eq.~\eqref{bitphaseerror2} (for $M=2$). For the $M>2$
case, we may increase the key generation rate by incorporating the
mutual information between bit and phase errors.
As in Ref.~\cite{FTL}, we may parameterize the diagonal elements as
follows: $p_x=e_b-\lambda$, $p_z=e_p-\lambda$, and $p_y=\lambda$.
The overall secret key generation rate  is
\begin{equation}
\label{eqn:keyrate2}
R_{\text{final}} = \frac{p_{con}}{M}
[1-H_4(1-e_b-e_p+\lambda,e_b-\lambda,\lambda,e_p-\lambda)].
\end{equation}
where $H_4(x_1,x_2,x_3,x_4) = \sum_{i=1}^{4} -x_i \log_2 x_i$.
The parameter $\lambda$ represents the mutual information, and in order to
prove security, one needs to find the worst-case value of $\lambda$ to compute
the worst-case key generation rate.
When one assumes pessimistically that $0\leq \lambda \leq e_b$, the worst-case
value can easily be shown to be $\lambda=e_b e_p$.
In this case, no mutual information, if any, is incorporated in
the key generation rate in
Eq.~\eqref{eqn:keyrate2} and it reduces to the rate in Eq.~\eqref{eqn:keyrate1}.
On the other hand,
for the case when
$M>2$,
we can see from Eqs.~\eqref{pi_calculations} to \eqref{pz_calculations} that
the admissible range of $\lambda$ is actually $e_b \cos^2 \theta \leq \lambda \leq e_b$.
In this case, the worst-case value of $\lambda$ is $\lambda=e_b \cos^2
\theta$ when $e_b < \cos^2 \theta/(1+\cos^2 \theta)$ and is $\lambda=e_b e_p$ otherwise.


As an example, we consider the SARG04 protocol ($M=4$, $\theta=\pi/4$).
Using Eq.~\eqref{bitphaseerror}, the relation between bit and phase error rates is
$e_p=3e_b/2$.  Assuming that $e_b < 1/3$, the worst-case value of $\lambda$ is $\lambda=e_b/2$.
These results agree with those in Ref.~\cite{TL,FTL}.

\section{Conclusions}

In summary, we have proposed and proved the security of a
generalized QKD protocol that includes the BB84 and SARG04 protocols
as specific cases.
 We have derived a closed form formula for the relationship between
 the bit and phase error rates of the generalized protocol.
 Such a relation
allows us to place bounds on the security of the protocol. Therefore
we have provided
 a unified security proof for protocols with discrete rotational symmetry.
 We observe that when the number of bases is larger than two, discrete symmetry simplifies the effect
of a noisy channel or eavesdropper, thus resulting in a simple
relationship between bit and phase error rates.
 In this
paper, we have only considered Abelian symmetry. In future, it may
be interesting to consider the role of non-Abelian discrete symmetry
in QKD. A similar security analysis involving a representation
theory summation (\ref{NewDensityMatrix}) may prove useful when
studying higher dimensional protocols such as the six-state
protocol.


\appendix

\section{Calculation of the components of $\rho_{AB}$ when $M>2$\label{app-caldiagonal}}

In this appendix, we calculate the diagonal elements of $\rho_{AB}$ given in Eq.~\eqref{AB_densitymatrix}.
Specifically, we demonstrate by example how to derive Eqs.~\eqref{pi_calculations}-\eqref{pz_calculations}
by explicitly deriving Eq.~\eqref{px_calculations}.
The other equations can be obtained in a similar way.
To proceed, we first prove the following lemma.


\begin{lemma}{\rm
The following are spherical averages
for $M > 2$:}
\begin{eqnarray}\label{lemma}
    \frac{1}{M}\sum_{l = 0}^{M-1} \cos^2 2l\pi/M &=& 1/2\\
    \frac{1}{M}\sum_{l = 0}^{M-1} \sin^2 2l\pi/M &=& 1/2\\
    \frac{1}{M}\sum_{l = 0}^{M-1} \cos 2l\pi/M \sin 2l\pi/M &=& 0
\end{eqnarray}
\end{lemma}
\begin{proof}

The trigonometric functions decompose into exponentials as:
\begin{eqnarray}\label{lemmaproof2}
    \cos^2 2l\pi/M &=& \frac{1}{4}(2 + e^{4l\pi/M} + e^{-4l\pi/M}) \phantom{xxx}\\
    \sin^2 2l\pi/M &=& \frac{1}{4}(2 +i e^{4l\pi/M} -i e^{-4l\pi/M}) \phantom{xxx} \\
    \cos 2l\pi/M \sin 2l\pi/M &=& -\frac{1}{2}i (e^{4l\pi/M} - e^{-4l\pi/M}) \phantom{xxxx}
\end{eqnarray}

In addition, we have the geometric series

\begin{eqnarray}\label{lemmaproof}
    \frac{1}{M}\sum_{l = 0}^{M-1} e^{\pm 4l\pi/M} &=& \frac{1-e^{\pm 4\pi}}{1 - e^{\pm 4\pi/M}}\\
    &=& 0
\end{eqnarray}
which follows directly for $M \neq 2$ since the denominator is finite.  Thus, substitution of
the former (\ref{lemmaproof2}) into the latter (\ref{lemmaproof}) yields the desired result.  In
the special case where $M = 2$, equation (\ref{lemmaproof}) is ill defined. Instead,
we have by inspection summations of the form: $\frac{1}{4}\sum_{l = 0}^{3} \cos^2 l\pi/2 = (1+1)/4
= 1/2$ and $\frac{1}{4}\sum_{l = 0}^{3} \sin^2 l\pi/2 = 1/2$, as required.
\end{proof}

\subsection{Calculation of $p_x$}

We first take the inner product of $\rho_{AB}$ given in Eq.~\eqref{AB_densitymatrix} with the Bell state corresponding to a bit flip error:
\begin{eqnarray}\label{px_calculation}
     p_x &=& 
     \langle\Psi^+|\rho_{AB}|\Psi^+\rangle\\
         &=& \frac{1}{M}\sum_{j}\sum_{l = 0}^{M-1} \Big|\langle\Psi^+| \widehat{I}_A \otimes \nonumber\\
         &&(\widehat{F}_0\widehat{R}_{-l\pi/M}\widehat{E}_j\widehat{R}_{l\pi/M})_B|\psi_0\rangle_{AB}\Big|^2
\end{eqnarray}

Temporarily dropping the summation over channel operators $\widehat{E}_j$ and inserting the x-basis
representations for $|\Psi^+\rangle$
and $|\psi\rangle_{AB}$ we obtain:
\begin{eqnarray}
        &=& \frac{1}{M}\sum_{l=0}^{M-1} \Big| \frac{1}{\sqrt{2}}(\langle00| - \langle11|)(\widehat{F}_0\widehat{R}_{-l\pi/M}\widehat{E}_j\widehat{R}_{l\pi/M})_B\times
                            \nonumber\\
                & & (\cos\frac{\theta}{2} |00\rangle_B + \sin\frac{\theta}{2}|11\rangle)\Big|^2\\
        &=& \frac{\sin^2\theta}{4M}\sum_{l=0}^{M-1} |\langle0|\widehat{R}_{-l\pi/M}\widehat{E}_j\widehat{R}_{l\pi/M}|0\rangle -\nonumber\\
	&&\langle1|\widehat{R}_{-l\pi/M}\widehat{E}_j\widehat{R}_{l\pi/M}|1\rangle|^2
\end{eqnarray}

Lastly, we insert our matrix representations for the operators $\widehat{R}_{l\pi/M}$ and
$\widehat{E}_{j}$:
\begin{eqnarray}
        &=& \frac{\sin^2\theta}{4M}\sum_{l=0}^{M-1} \Big|a_x\cos2l\pi/M - a_z\sin2l\pi/M\Big|^2\\
        &=& \frac{\sin^2\theta}{4M}\sum_{l=0}^{M-1} \Big(|a_x|^2\cos^2 2l\pi/M + |a_z|^2\sin^2 2l\pi/M -
                            \nonumber\\
                & & (a_xa_z^*+a_za_x^*)\cos 2l\pi/M\sin 2l\pi/M\Big)
\end{eqnarray}
which, by our preliminary lemma, reduces to:
\begin{equation}
        p_x = \frac{1}{8} \sin^2\theta (|a_x|^2+|a_z|^2).
\end{equation}
By reintroducing the summation over $j$ for the channel operators $\widehat{E}_j$ and normalizing, we obtain Eq.~\eqref{px_calculations}.
The normalization constant in Eq.~\eqref{px_calculations} is
$N'=2 \sum_j |a_i^j|^2\sin^2\theta + |a_x^j|^2 + (1+\cos^2\theta)|a_y^j|^2+ |a_z^j|^2$.
The expressions for $p_i$, $p_y$, and $p_z$ can also be obtained in a similar way.

\section{Relation between bit and phase error rates when $M=2$\label{app-epebrelation}}

In this appendix, we find the relation between the bit and phase error rates for the $M=2$ case
by solving
the problem of maximizing the phase error rate given in Eq.~\eqref{eqn-M2ep}
subject to a fixed bit error rate given in Eq.~\eqref{eqn-M2eb}.
This problem can be written as follows:
\begin{eqnarray}
&&\hspace{-1.3cm}\text{maximize} \nonumber \\
\label{eqn-app2-p1-1}
     e_p &=& (|a_y|^2 + |a_z|^2)(1 + \cos^2\theta) \\
&&\hspace{-1.3cm}\text{subject to} \nonumber \\
\label{eqn-app2-p1-2}
     e_b &=& |a_x|^2\sin^2\theta + |a_y|^2 + |a_z|^2 \cos^2\theta \\
\label{eqn-app2-p1-3}
     1   &=& (|a_i|^2 + |a_x|^2)\sin^2\theta + \nonumber \\
         & & (|a_y|^2 + |a_z|^2)(1 + \cos^2\theta) \\
\label{eqn-app2-p1-4}
    && \hspace{-.9cm}|a_i|^2 , |a_x|^2,|a_y|^2 , |a_z|^2 \geq 0
\end{eqnarray}
where the maximization is over $|a_\beta|^2 \triangleq \frac{1}{N''} \sum_j |a_\beta^j|^2$, $\beta \in \{i,x,y,z\}$.
Here, the second constraint in Eq.~\eqref{eqn-app2-p1-3} is due the normalization requirement that
$p_i+ p_x+p_y+p_z=1$.
Since there are two equality constraints and four optimization variables, we may eliminate two variables,
namely $|a_y|^2$ and  $|a_z|^2$, and simplify the problem to
\begin{eqnarray}
&&\hspace{-1.1cm}\text{maximize} \nonumber \\
\label{eqn-app2-p2-1}
     &&e_p = 1-(|a_i|^2 + |a_x|^2)\sin^2\theta \\
&&\hspace{-1.1cm}\text{subject to} \nonumber \\
\label{eqn-app2-p2-2}
&&\frac{1}{\sin^2\theta }
\begin{bmatrix}
e_b - A\\
-e_b+B
\end{bmatrix}
+
\begin{bmatrix}
A & -B\\
-B & A
\end{bmatrix}
\begin{bmatrix}
|a_i|^2\\
|a_x|^2
\end{bmatrix}
\geq 0 \\
    && |a_i|^2 , |a_x|^2 \geq 0
\end{eqnarray}
where $B=1/(1+\cos^2\theta)$ and $A=B \cos^2\theta$.
Here, the objective function in Eq.~\eqref{eqn-app2-p2-1} is obtained by substituting Eq.~\eqref{eqn-app2-p1-3} in Eq.~\eqref{eqn-app2-p1-1};
the left-hand side of Eq.~\eqref{eqn-app2-p2-2} is obtained by solving Eqs.~\eqref{eqn-app2-p1-2} and \eqref{eqn-app2-p1-3} for $[|a_y|^2;|a_z|^2]$, which are both greater than zero according to
Eq.~\eqref{eqn-app2-p1-4}, thus leading to the inequality in Eq.~\eqref{eqn-app2-p2-2}.
Note that in general $|a_i|^2 = |a_x|^2=0$ (giving arise to $e_p=1$) may not be a feasible solution.
The simplified problem is a linear programming problem;
the four constraints in the problem correspond to four lines and the objective function corresponds to a line in a two-dimensional space.
Suppose that we are interested in the case when $e_b \leq A$ (which turns out to cover the entire range of $e_p$, $0 \leq e_p \leq 1$).
Given that
$e_b \leq A$, one can easily find the feasible region and the maximizing point $(|a_i|^2 , |a_x|^2)$.
It turns out that the maximizing point is
\begin{equation}
(|a_i|^2 , |a_x|^2)=\left(\frac{A-e_b}{A \sin^2\theta}, 0\right) ,
\end{equation}
and the maximum phase error rate is
\begin{equation}
     e_p = \frac{1+\cos^2 \theta}{\cos^2 \theta} e_b, \phantom{xxx} \text{for } M=2 \text{ and } e_b \leq \frac{\cos^2 \theta}{1+\cos^2 \theta}. 
\end{equation}


\begin{thebibliography}{99}
\bibitem {NoteB} Our summation for $\Phi$ echoes Schur's lemma, however we work with reducible
representations of the rotation operators. Hence, $\Phi$ is not proportional to the identity.
\bibitem {NoteA} Note that one can show if $e_b = 0$, the normalized conclusive probability
$p_{con}=\frac{1}{2}\sin^2 \theta$ while $e_b>0$ implies $p_{con}>\frac{1}{2}\sin^2 \theta$.
\end{thebibliography}
\end{document}